\documentclass[12pt]{article}

\usepackage{epsf}
\usepackage{epsfig}
\usepackage{cite}
\usepackage{amsmath}
\usepackage{graphics}

\begin{document}

\setlength{\textheight}{21.5cm}
\setlength{\oddsidemargin}{0.cm}
\setlength{\evensidemargin}{0.cm}
\setlength{\topmargin}{0.cm}
\setlength{\footskip}{1cm}
\setlength{\arraycolsep}{2pt}

\renewcommand{\thefootnote}{\#\arabic{footnote}}
\setcounter{footnote}{0}

\newcommand{\gtrsim}{ \mathop{}_{\textstyle \sim}^{\textstyle >} }
\newcommand{\lesssim}{ \mathop{}_{\textstyle \sim}^{\textstyle <} }
\newcommand{\rem}[1]{{\bf #1}}
\renewcommand{\thefootnote}{\fnsymbol{footnote}}
\setcounter{footnote}{0}
\def\thefootnote{\fnsymbol{footnote}}

\hfill {\tt arXiv:08mm.nnnn[hep-ph]}\\
\vskip .5in

\begin{center}

\bigskip
\bigskip

{\Large \bf Predictions of Neutrino Mixing Angles in a 
$T^{'}$ Model}

\vskip .45in

{\bf David A. Eby\footnote{daeby@physics.unc.edu},
Paul H. Frampton\footnote{frampton@physics.unc.edu}
and Shinya Matsuzaki\footnote{synya@physics.unc.edu}} 

\vskip .3in

{\it Department of Physics and Astronomy, University of North Carolina,
Chapel Hill, NC 27599-3255.}

\end{center}

\vskip .4in 
\begin{abstract}
Flavor symmetry 
($T^{'} \times Z_2$) where
$T^{'}$ is the binary tetrahedral group
predicts for neutrino mixing angles
$\theta_{13} = \sqrt{2} \left( \frac{\pi}{4} - \theta_{23} \right)$
and, with one phenomenological input, provides
upper and lower bounds on both 
$\theta_{13}$ and 
$\theta_{23}$. 
The predictions arise from
the deviation
of the Cabibbo angle $\Theta_{12}$
from its lowest-order 
value $\tan 2\Theta_{12} = (\sqrt{2})/3$
and from the $T^{'}$ mechanism which relates
mixing of $(\nu_{\tau}, \nu_{\mu}, \nu_e)$
neutrinos to mixing of $(s, d)$ quarks.
\end{abstract}

\renewcommand{\thepage}{\arabic{page}}
\setcounter{page}{1}
\renewcommand{\thefootnote}{\#\arabic{footnote}}

\newpage

As an attractive alternate to the grand unification of 
the strong and electroweak interactions, 
a global flavor symmetry acting in tandem with the
standard model gauge group can address the issue
of relating the quarks and leptons. Other than the
equality of numbers of quarks and leptons
no direct connection between them has
been discovered.

\bigskip

Evidence for grand unification such as proton decay
remains elusive so it seems worth finding testable
predictions of family symmetry. 

\bigskip

In the present article we shall invoke a family
symmetry involving the binary tetrahedral group $T^{'}$
which has sufficient structure to relate quarks
and leptons.
In a previous study\cite{Tprime},
an exact formula for the Cabibbo angle $\Theta_{12}$
was derived in a $(T^{'} \times Z_2)$ model where the neutrino
mixing angles are of the tribimaximal (TBM)
values\cite{HPS}.
In the same model, a striking prediction was made\cite{FMHiggs}
for two-body leptonic decays of the
Higgs boson.

\bigskip

The three neutrino mixing angles $\theta_{12}, \theta_{23}, \theta_{13}$
are empirically consistent with the TBM values. However,
as the experimental accuracy improves, this situation may change.
Thus, it is of considerable interest to predict quantitatively
what departures from the TBM values
\begin{equation}
\theta_{12} = \tan^{-1} \left( \frac{1}{\sqrt{2}} \right),
~~~ \theta_{23} = \left( \pi/4 \right), ~~~ \theta_{13} =0
\label{TBM}
\end{equation}
are to be expected? We are delighted to report that the $(T^{'} \times Z_2)$
model allows one to address this question by relating
the perturbations around TBM
\begin{equation}
\theta_{ij} = (\theta_{ij})_{TBM} + \epsilon_k,
\label{PerturbedTBM}
\end{equation}
where we use $\epsilon_3$ for $\theta_{12}$, and so on,
and the TBM values
are in Eq.(\ref{TBM}), to the perturbation around the Cabibbo
angle value
\begin{equation}
\tan 2(\Theta_{12})_{T^{'}}  = \left( \frac{1}{3} (\sqrt{2}) \right).
\end{equation}
\label{CabibboTprime}

\bigskip

We are less delighted with progress on the quark and lepton
masses. Although we understand 
why $m_t \gg m_b > m_{c,s,d,u}$ for quarks and why $m_3 \gg m_{1,2}$
for neutrinos, when we look more closely at the details
we find that masses are not quantitatively explained.
It is not clear to us whether this will
be corrected in the $(T^{'} \times Z_2)$ model
by higher order corrections and/or adding $T^{'}$ doublet
VEVs.
We hope to return to the masses in a further work.

\bigskip

In the present work, we take the view that
the model can make reliable predictions about mixing angles
even when details of the mass spectra are incomplete.

\bigskip

To analyze the relationship between the perturbations
in Eq.(\ref{PerturbedTBM}) and the Cabibbo angle
will require, as we shall see, very interesting $T^{'}$
algebra sometimes arriving at astonishingly simple expressions.

\bigskip

Let us begin the analysis. 

\bigskip

First we recall a few salient points about the model
in \cite{FMA4} based on $A_4$ symmetry\cite{Ma,Ma2,Altarelli}. 
The only important scalar for the
present analysis is the triplet $H_3(3, +1)$
whose vacuum expectaion value in \cite{FMA4} was taken as
\begin{equation}
<H_3> = (V_1, V_2, V_3) = V(1, -2, 1)
\label{H3TBM}
\end{equation}
which led to the TBM neutrino mixing in Eq.(\ref{TBM}).
We consider the perturbation 
\begin{equation}
<H_3> = (V^{'}_1, V^{'}_2, V^{'}_3) = V^{'}(1, -2+b, 1+a)
\label{H3new}
\end{equation}
where $|a|, |b| \ll 1$.

We first consider the calculation which makes the
perturbation around the $\Theta_{12}$ calculation
based on $T^{'}$ symmetry \cite{Kephart0,Carr,Feruglio,mahanthappa}
in \cite{Tprime} by using Eq.(\ref{H3new})
in place of Eq.(\ref{H3TBM}). 
The down-quark $(2 \times 2)$ mass matrix for the first
two families $(s, d)$ is perturbed to
\begin{equation}
D \equiv \left( \frac{1}{V^{'} Y_{{\cal S}}} \right)
D^{'}  = \left( \begin{array}{cc}
\frac{1}{\sqrt{3}} & (-2+b) \sqrt{\frac{2}{3}} \omega \\
\sqrt{\frac{2}{3}}(1+a) & \frac{1}{\sqrt{3}} \omega
\end{array}
\right)
\label{Dmatrix}
\end{equation}

\bigskip

The hermitian square 
${\bf {\cal D}} \equiv D D^{\dagger}$
is 
\begin{equation}
{\bf {\cal D}} \equiv D D^{\dagger} \simeq \left(
\frac{1}{3} \right) \left(
\begin{array}{cc}  9 - 8b & \sqrt{2} (-1+a+b)  \\
\sqrt{2} (-1+a+b) & 3+4a 
\end{array}
\right),
\label{DDdagger}
\end{equation}
The eigenvalues satisfy the quadratic equation
\begin{equation}
(9 - 8b-\lambda)(3+4a-\lambda) - 2 (1-a-b)^2 = 0
\label{quadratic}
\end{equation}
with solutions
\begin{equation}
\lambda_{\pm} = (6 \pm \sqrt{11}) + 2a\left( 1 \mp \frac{4}{\sqrt{11}} \right)
-2b \left( 2 \pm \frac{7}{\sqrt{11}} \right)
\label{eigenvalues}
\end{equation}
An eigenvector $(\alpha, \beta)$  has components satisfying
\begin{equation}
\left( \frac{\beta}{\alpha} \right) = \left(
\frac{3 - \sqrt{11}}{\sqrt{2}} \right) \left[ 1 - \frac{a}{\sqrt{11}}
+ \frac{b}{\sqrt{11}} \right]
\label{components}
\end{equation}
whose normalization $N(\alpha, \beta)$ satisfies
\begin{equation}
N^{-2} = 1 + \beta^2/\alpha^2
\label{normalization}
\end{equation}
from which the Cabibbo angle $\sin \Theta_{12} = N \beta/\alpha$ is
\begin{equation}
\sin \Theta_{12} = \sqrt{ \left(\frac{1}{2} - \frac{3}{2 \sqrt{11}} \right) }
\left(1 - \frac{3 + \sqrt{11}}{22} (a - b) \right)
\label{sin}
\end{equation}

\bigskip

\noindent From this one finds at leading order 
\begin{equation}
\cos 2 \Theta_{12} \simeq \left(\frac{3}{\sqrt{11}} \right)
\left( 1+ \frac{2}{33} (a - b) \right)
\label{Cos}
\end{equation}
and
\begin{equation}
\sin 2\Theta_{12} \simeq \left(\frac{\sqrt{2}}{\sqrt{11}} \right) 
\left( 1 - \frac{3}{11} (a - b) \right),
\label{Sin}
\end{equation}
whence
\begin{equation}
\tan 2 \Theta_{12} \simeq (\sqrt{2})/3  \left(1 - \frac{1}{3} (a - b) \right)
\label{Tan}
\end{equation}

\bigskip

\noindent which is a suprisingly simple generalization
of the $a=b=0$ case \cite{Tprime}!

\bigskip

Our next step is to relate the perturbations $\epsilon_i$
in the neutrino mixing angles, Eq.(\ref{PerturbedTBM}), to the
perturbations $a$ and $b$ in the vacuum alignment, Eq.(\ref{H3new}).

\bigskip

We use a neutrino mixing matrix relating flavor eigenstates to
mass eigensates $(\nu_1, \nu_2, \nu_3)^T \equiv U (\nu_{\tau}, \nu_{\mu}, \nu_e)^T$
with, assuming no CP violation
\begin{equation}
U = \left( \begin{array}{ccc}
-s_{12}s_{23} - c_{12}c_{23}s_{13} &  - s_{12} c_{23} + c_{12}s_{23}s_{13} & + c_{12}c_{13} \\
+ c_{12}s_{23} - s_{12}c_{23}s_{13} & + c_{12}c_{23} +s_{12}s_{23}s_{13} & + s_{12}c_{13} \\
+ c_{23}c_{13} & + s_{23}c_{13} & +s_{13} 
\end{array}
\right).
\label{PMNS}
\end{equation}
This takes the TBM form 

\begin{equation}
U_{TBM}  = \left( \begin{array}{ccc}
-\sqrt{\frac{1}{6}} & - \sqrt{\frac{1}{6}} & + \sqrt{\frac{2}{3}}
\\
+ \sqrt{\frac{1}{3}} & + \sqrt{ \frac{1}{3}} & + \sqrt{\frac{1}{3}} \\
+ \sqrt{\frac{1}{2}} & - \sqrt{\frac{1}{2}} & 0
\end{array}
\right)\label{Utbm}
\end{equation}
for the values of the neutrino mixing angles $\theta_{12}, \theta_{23},
\theta_{13}$ given in Eq.(\ref{TBM}).

\bigskip

\noindent Making the perturbations defined in Eq.(\ref{PerturbedTBM}), one has, at first order,

\bigskip

$s_{12} \simeq \sqrt{\frac{1}{3}}(1+\sqrt{2}\epsilon_3); ~~ 
c_{12} \simeq \sqrt{\frac{2}{3}}(1-\epsilon_3/\sqrt{2});$

\bigskip

$s_{23} \simeq \sqrt{\frac{1}{2}}(1+\epsilon_1); ~~ c_{23} \simeq \sqrt{\frac{1}{2}}(1-\epsilon_1)$.

\bigskip

$s_{13} \simeq \epsilon_2; ~~ c_{13} \simeq  1$;

\bigskip

\noindent Consequently one may write
\begin{equation}
U \simeq U_{TBM} + \delta U = U_{TBM} + 
\delta U_1\epsilon_1 + \delta U_2 \epsilon_2 + \delta U_3 \epsilon_3
\label{deltaU}
\end{equation}
in which

\bigskip

\begin{equation}
\delta U_1 = \left( \begin{array}{ccc}
- \sqrt{\frac{1}{6}} & + \sqrt{\frac{1}{6}} & 0 \\
+ \sqrt{\frac{1}{3}} & -\sqrt{\frac{1}{3}} & 0 \\
- \sqrt{\frac{1}{2}} & - \sqrt{\frac{1}{2}} & 0
\end{array}
\right)
\label{deltaU1}
\end{equation}

\begin{equation}
\delta U_2 = \left( \begin{array}{ccc}
- \sqrt{\frac{1}{3}} & + \sqrt{\frac{1}{3}} & 0 \\
- \sqrt{\frac{1}{6}} & + \sqrt{\frac{1}{6}} & 0 \\
0 & 0 & 0
\end{array}
\right)
\label{deltaU2}
\end{equation}

\begin{equation}
\delta U_3 = \left( \begin{array}{ccc}
- \sqrt{\frac{1}{3}} & - \sqrt{\frac{1}{3}} & - \sqrt{\frac{1}{3}} \\
- \sqrt{\frac{1}{6}} & -\sqrt{\frac{1}{6}} & + \sqrt{\frac{2}{3}} \\
0 & 0 & 0
\end{array}
\right)
\label{deltaU3}
\end{equation}

For TBM mixing, one has
\begin{equation}
(M_{\nu})_{TBM} = U_{TBM}^T (M_{\nu})_{diag} U_{TBM} 
~~~ {\rm with} ~~~ (M_{\nu})_{diag} = \left( \begin{array}{ccc}
m_1 & 0 & 0 \\
0 & m_2 & 0 \\
0 & 0 & m_3 
\end{array}
\right)
\label{Mdiag}
\end{equation}

\bigskip

\noindent This gives

\begin{equation}
(M_{\nu})_{TBM}
= 
\left( \frac{1}{6} \right) 
\left( \begin{array}{ccc}
m_1 + 2m_2 + 3m_3 & m_1 + 2m_2 -3m_3 & -2m_{12} \\
& m_1 + 2m_2 + 3 m_3 & -2 m_{12} \\
&  &  4 m_1 + 2m_2 
\end{array}
\right)
\label{MnuTBM}
\end{equation}
where $(M_{\nu})_{TBM}$ is symmetric and where $m_{12} \equiv (m_1 - m_2)$.

\newpage

\noindent From Eq.(\ref{Mdiag}) the perturbation in $(M_{\nu})_{diag}$
satisfies
\begin{eqnarray}
\delta (M_{\nu})_{diag} & = &  \left( \begin{array}{ccc}
\delta m_1 & 0 & 0 \\
0 & \delta m_2 & 0 \\
0 & 0 & \delta m_3 
\end{array}
\right)  \nonumber \\
& = &  \delta U (M_{\nu})_{TBM} U_{TBM}^T \nonumber \\
&+&  U_{TBM} \delta M_{\nu} U_{TBM}^T \nonumber \\
&+& U_{TMB} (M_{\nu})_{TBM} \delta U^T,
\label{deltaMdiag}
\end{eqnarray}
in which $U_{TBM}$ is known from Eq.(\ref{Utbm})
and $\delta U$ from Eqs. (\ref{deltaU},\ref{deltaU1},\ref{deltaU2},\ref{deltaU3}). 

\bigskip

To compute $\delta M_{\nu}$ in Eq.(\ref{deltaMdiag})
we use $(M_{\nu})_{TBM}$ from reference \cite{FMA4}
\begin{equation}
(M_{\nu})_{TBM} =
\left(
\begin{array}{ccc}
x_1 V_1^2 + 2 x_{23} V_2V_3 &
~~ x_1V_1V_3 + x_{23} (V_2^2+V_1V_3) &
~~ x_1V_1V_2 + x_{23} (V_3^2+V_1V_2) \\
& 
~~ x_1V_3^2 + 2 x_{23} V_1V_2 & 
~~ x_1V_2V_3 + x_{23} (V_1^2+V_2V_3)  \\
&
& 
~~ x_1V_2^2 + 2 x_{23} V_1V_3
\end{array}
\right).
\label{MnuA4}
\end{equation}

\bigskip

\noindent in which $<H_3> = (V_1, V_2, V_3)$, $x_1=Y_1^2/M_1$
and $x_{23} = Y_2Y_3/M_{23}$. These variables
involve Yukawa couplings and right-handed neutrino masses  
all of which are empirically unknown.
Only the combination $y=x_{23}/x_1$ survives and 
our predictions will be obtained by eliminating this unknown.

\bigskip

To find $\delta M_{\nu}$ in Eq.(\ref{deltaU})
we use the perturbation of the vacuum alignment, Eq.(\ref{H3new}),
in Eq.(\ref{MnuA4}) to find 
\begin{equation}
\delta M_{\nu} =
V_1^{'2} x_1 \left(
\begin{array}{ccc}
~~ 2(-2a+b)y & a + (a-4b)y & b+(2a+b)y \\
 & 2(a+by) & (-2a+b)(1+y) \\
  &  &  -4b +2ay 
\end{array}
\right).
\label{MA4}
\end{equation}

\bigskip

By inserting this $\delta M_{\nu}$ 
into Eq.(\ref{deltaMdiag}) we obtain
six equations from the $(3 \times 3)$ symmetric
matrix to combine with Eq.(\ref{Tan}) above.
In the $\delta m_1$ of (I) - (III) a common (unpredicted)
normalization factor has been omitted.

\begin{itemize}

\item (I) $\delta m_1 = (2+y) (a -2b)$

\item (II) $\delta m_2 = 0$

\item (III) $\delta m_3 = -3y (a-2b)$

\item (IV)  $\epsilon_2 = - \sqrt{2} \epsilon_1$

\item (V)  $a = 6 \epsilon_1 = -3 \sqrt{2} \epsilon_2$

\item (VI) $(a+b)  = \left( \frac{3}{\sqrt{2}} \frac{2+y}{1-y} \right)\epsilon_3$

\end{itemize}

\newpage

The result (IV) provides a prediction from $T^{'}$ that
\begin{equation}
\theta_{13} = \sqrt{2} \left( \frac{\pi}{4} - \theta_{23} \right)
\label{root2}
\end{equation}
which interestingly links any non-zero value for $\theta_{13}$
to the departure of the atmospheric neutrino mixing angle
$\theta_{23}$ from maximal mixing with $\theta_{23} = \pi/4$.
This is our most definite prediction from $T^{'}$,
independent of phenomenoligical input
\footnote{A similar prediction from a different
starting point appeared in \cite{HS}.}.

\bigskip

To arrive at further $T^{'}$ predictions for the neutrino
mixings $\theta_{13}$ and $\theta_{23}$ we shall require
phenomenological input.

\bigskip

The equation (I) through (III) must be combined with the zeroth
order values
\begin{equation}
m_1^0 = 3 (y+2),
\label{m10}
\end{equation}
\begin{equation}
m_2^0 = 0, 
\label{m20}
\end{equation}
\begin{equation}
m_3^0 = - 9y. 
\label{m30}
\end{equation}

\bigskip

\noindent It is noted that $m_2 = 0$ remains even at 
first order. This arises from the zero structures
\cite{textures,marfatia,FGY} in the terms
of Eq.(\ref{deltaMdiag}). They are
\begin{equation}
\delta U (M_{\nu})_{TBM} U_{TBM}^T 
~~~~~~~~~~ \left(\begin{array}{ccc}
0 & 0 & \\
& 0 & \\
& 0 & 0
\end{array}
\right),
\label{term1}
\end{equation}
\begin{equation}
U_{TBM} \delta M_{\nu} U_{TBM}^T 
~~~~~~~~~~ \left(\begin{array}{ccc}
& & \\
& 0 & 0 \\
 & 0 & 
\end{array}
\right),
\label{term2}
\end{equation}
\begin{equation}
U_{TBM} (M_{\nu})_{TBM} \delta U^T 
~~~~~~~~~~ \left(\begin{array}{ccc}
0 &  &  \\
0 & 0 & 0 \\
 & & 0
\end{array}
\right).
\label{term3}
\end{equation}

\bigskip

\noindent The necessary phenomenological input,
exactly as in \cite{FMA4}, is to set 
\begin{equation}
y=-2 
\label{y=-2}
\end{equation}
from which equation (VI) gives $(a+b)=0$ and Eq.(\ref{Tan}) becomes
simply
\begin{equation}
\tan 2 \Theta_{12} = \left( \frac{\sqrt{2}}{3} \right) \left(1 - 4 \epsilon_1 \right)
\label{TanNew}
\end{equation}

\bigskip

\noindent Eq.(\ref{TanNew}) allows us, from the 
experimental value \cite{PDG2008},
$(\Theta_{12})_{experiment} = 13.05 \pm 0.07^o$,
to identify the limits
\begin{equation}
- 0.0114 < \epsilon_1 < - 0.0082
\label{eps1}
\end{equation}
and 
\begin{equation}
0.011 <  \epsilon_2 < 0.016 
\label{eps2}
\end{equation}

\bigskip

\noindent The values in Eqs.(\ref{eps1},\ref{eps2})
of $\epsilon_{1,2}$ lead directly
to predictions for the neutrino mixing angles.
Substitution of Eqs.(\ref{eps1},\ref{eps2})
into Eq.(\ref{PerturbedTBM}) gives

\bigskip

\begin{equation}
0.5 \times 10^{-3} \leq \sin^2 2 \theta_{13} \leq 1.0 \times 10^{-3}
\label{predict13}
\end{equation}
and
\begin{equation}
0.99947 \leq \sin^2 2 \theta_{23} \leq 0.99973
\label{predict23}
\end{equation}

\bigskip

The situation with respect to $T^{'}$ flavor symmetry
is very exciting. The predictions Eq. (\ref{root2}),
Eq. (\ref{predict13}) and Eq. (\ref{predict23}) have different
status. The prediction relating $\theta_{13}$
and $\theta_{23}$ in Eq.(\ref{root2}) is the
sharpest.
With the one phenomenological input, Eq.(\ref{y=-2}),
necessary to obtain a sensible neutrino mass spectrum
one arrives at the predictions in Eq.(\ref{predict13})
and Eq.(\ref{predict23}) which also provide targets of
opportunity
for experiments.

\bigskip

\noindent With regard to quark and lepton masses, the flavor
symmetry leaves them as enigmatic as before, basically
as free parameters just as for the standard model.
The mixing angles are, however, significantly
constrained by the $T^{'}$ geometrical
structure. In particular, we have shown how the neutrino
mixing angles are predicted by the empirical
departure from the $T^{'}$ prediction
for the largest quark mixing of the Cabibbo angle.

\newpage

\begin{center}

\section*{Acknowledgements}

\end{center}

This work was supported in part 
by the U.S. Department of Energy under Grant
No. DE-FG02-06ER41418.

\bigskip
\bigskip
\bigskip
\bigskip

\end{document}